\documentstyle[aps,psfig]{revtex} 
\newcommand{\be}{\begin{equation}} 
\newcommand{\ee}{\end{equation}} 
\newcommand{\ben}{\begin{eqnarray}} 
\newcommand{\een}{\end{eqnarray}}    
\begin{document} 
 
\twocolumn[\hsize\textwidth\columnwidth\hsize\csname 
@twocolumnfalse\endcsname 
 
\title{Bags, junctions, and networks of BPS and non-BPS defects} 
\author{D. Bazeia and F.A. Brito}
\address{Departamento de F\'\i sica, Universidade Federal da Para\'\i ba,\\ 
Caixa Postal 5008, 58051-970 Jo\~ao Pessoa, Para\'\i ba, Brazil} 
 
\date{\today} 

\maketitle

\begin{abstract}

We investigate several models of coupled scalar fields that present
discrete $Z_2, Z_2\times Z_2$, $Z_3$ and other symmetries. These
models support topological domain wall solutions of the BPS and non-BPS
type. The BPS solutions are stable, but the stability of the
non-BPS solutions may depend on the parameters that specify
the models. The BPS and non-BPS states give rise to bags, and also
to three-junctions that may allow the presence of networks of topological
defects. In particular, we show that the non-BPS defects
of a specific model that engenders the $Z_3$ symmetry give rise
to a stable regular hexagonal network of domain walls.
\bigskip

\end{abstract}
\vskip2pt] 

\section{Introduction} 
 
Bogomol'nyi, and Prasad and Sommerfield \cite{bps75}, have presented
a way to find topological solutions that saturate the lower bound in energy.
These solutions are known as BPS states, and they are supposed to play some
role in the development of modern particle physics, because there are
supersymmetric models where they are shorter multiplets, stable,
and no continuum variation of parameters can modify these BPS states.
In Refs.~{\cite{bds95,bsa96,brs96e}} one has investigated some models 
of coupled real scalar fields in bidimensional space-time. These models 
present peculiar features, such as the presence of the Bogomol'nyi bound, 
with the corresponding appearance of first-order differential equations 
that solve the equations of motion in the case of static configurations. 
These Refs.~{\cite{bds95,bsa96,brs96e}} have found topological
solitons in specific models by taking advantage of the trial-orbit
method first introduced by Rajaraman in Ref.~{\cite{raj79}}. 
These investigations provide a concrete way of finding BPS states and 
suggest several other studies, in particular on the subject of defects 
inside defects, as presented for instance in Refs.~{\cite{wit85,lsh85,mke88}} 
and more recently in \cite{mor95,mor95b,brs96,bba97,mor98,etb98,bbb99}.  
The investigations of defects inside defects presented in 
Ref.~{\cite{brs96,bba97,etb98,bbb99}} are natural extensions
of the former papers \cite{bds95,bsa96,brs96e}. 
 
In supersymmetric models with chiral superfields the presence of 
discrete symmetry may produce BPS and non-BPS walls. The non-BPS
walls are solutions of the equations of motion, and they do not
necessarily saturate the lower bound in energy. But the BPS walls
saturate the lower bound in energy, and lie in shorter multiplets,
in a way such that they preserve the supersymmetry only partially
\cite{wol78,agi89,dsh97}. There are BPS states that preserve $1/2$ and $1/4$
of the supersymmetry \cite{wol78,agi89,dsh97,cqr91,ato91,gto99,cht99,saf99},
and in the present work we investigate several explicit examples.
We start by first offering general considerations in
Sec.~{\ref{sec:gmodels}}. We examine specific models in
Sec.~{\ref{sec:models}}, where we find several solutions, of the 
BPS and non-BPS type. See also
the more recent work \cite{oin99,cht99a}, where one can find other
investigations concerning the BPS domain wall junctions in supersymmetric
models.

Our work continues in Sec.~{\ref{sec:stability}}, where we investigate
stability of several BPS and non-BPS solutions. This is done not only to show
stability, but also to see how the bosonic matter behaves in the background
of BPS and non-BPS defects. The behavior of the fermionic matter in the
extended supersymmetric models can be investigated by following the lines
of Refs.~{\cite{etb98,mba96,csh97}}. In Sec.~{\ref{sec:further}} we
investigate two other issues. The first in Sec.~{\ref{sec:bag}},
where we consider the entrapment of the other field, as in bag
models \cite{cjj74,rwe75,mon76}. This possibility may be
implemented in the first, second and third models, and the bag
may be of the BPS or non-BPS type, depending of the particular 
model under consideration. We examine the second issue in
Sec.~{\ref{sec:junctions}}, where we investigate fusion of defects.
Here we briefly review Ref.~{\cite{ato91}}, which introduces investigations
valid within the supersymmetric context, and after we offer
another possibility of showing fusion of defects and stability
of junctions, valid beyond the context of supersymmetry.

In Sec.~{\ref{sec:network}} we show how one can generate
planar networks of BPS and non-BPS defects. There we show that the
three-junction that appears in supersymmetric models may be stable,
but cannot generate a BPS network of defects. However, when we give up
supersymmetry we find a model where domain wall junctions are stable,
the three-junction strictly obey the triangle inequality and generate
a stable network of defects. We summarize the main results in
Sec.~{\ref{sec:comments}}, where we end the present work.

The subject of this work may find direct applications in diverse
branches of physics. Besides the recent interest, put forward in
Refs.~{\cite{gto99,cht99,saf99}}, we recall for instance
that the $Z_3$ group is the center of $SU(3)$, and this may
guide us toward applications involving strong interactions. Some examples
can be found in Ref.~{\cite{ctr98,chw98}}. Another possibility is related
to the $Z_2\times Z_3$ symmetry, in an effort to entrap the $Z_3$-symmetric
portion of the model inside the topological defect generated by the $Z_2$
symmetry. Other applications include issues related to cosmology \cite{vsh94}
and magnetic materials \cite{esc81}, and to pattern formation in systems of
condensed matter \cite{wal97}. The investigations that we present are done
in flat space-time, in the $(d,1)$ dimensional case. We use natural units,
taking $\hbar=c=1$, and the metric tensor presents signature in the form
$(+,-,...,-)$. In the present work we are mainly concerned with the
cases $d=1,2,$ and $3$. We specify the spatial dimension along
with the interest of the subject under investigation.

\section{General considerations} 
\label{sec:gmodels} 
 
Let us start with two real scalar fields $\phi$ 
and $\chi$. We consider models defined via the Lagrangian density 
\be 
\label{L} 
{\cal L}=\frac{1}{2}\partial_{\alpha}\phi\partial^{\alpha}\phi+ 
\frac{1}{2}\partial_{\alpha}\chi\partial^{\alpha}\chi-V 
\ee 
This is valid in $(d,1)$ space-time dimensions. $V=V(\phi,\chi)$
is the potential. It has the general form 
\be 
\label{ssys} 
V(\phi,\chi)=\frac{1}{2}\,W^2_{\phi}+\frac{1}{2}\,W^2_{\chi} 
\ee 
$W=W(\phi,\chi)$ is some smooth function of the fields 
$\phi$ and $\chi$, and $W_{\phi}=(\partial W/\partial\phi)$ etc. 
This is the form of the (real bosonic portions of the) 
models we deal with in the present work. These models admit 
a supersymmetric extension, and in the supersymmetric theory the 
function $W$ is the superpotential. 
 
Models defined via $W(\phi,\chi)$ present several interesting 
properties \cite{bds95,bsa96,brs96e}. The equations of motion are  
\ben 
\label{t1} 
\frac{\partial^2\phi}{\partial t^2}-\nabla^2\phi+ 
\frac{\partial V}{\partial\phi}&=&0 
\\ 
\label{t2} 
\frac{\partial^2\chi}{\partial t^2}-\nabla^2\chi+ 
\frac{\partial V}{\partial\chi}&=&0 
\een 
For static field configurations living in the $(1,1)$ space-time, 
that is, for fields $\phi=\phi(x)$ and $\chi=\chi(x)$ they reduce to 
\ben 
\label{x1} 
\frac{d^2\phi}{dx^2}&=&W_{\phi}W_{\phi\phi}+W_{\chi}W_{\chi\phi} 
\\ 
\label{x2} 
\frac{d^2\chi}{dx^2}&=&W_{\phi}W_{\phi\chi}+W_{\chi}W_{\chi\chi} 
\een 
These equations are solved by first-order differential equations 
\ben 
\label{b1} 
\frac{d\phi}{dx}&=&W_{\phi} 
\\ 
\label{b2} 
\frac{d\chi}{dx}&=&W_{\chi} 
\een  
The energy of solutions that solve the first-order equations are bounded 
to the value $E^{ij}_B=|\Delta W_{ij}|$, where 
\be 
\Delta W_{ij}=W(\phi_i,\chi_i)-W(\phi_j,\chi_j) 
\ee 
Here $(\phi_i,\chi_i)$ is the i-th vacuum state of the model under
consideration. The reason for this is simple: the energy of static
configurations is
\be 
\label{e} 
E=\frac{1}{2}\int^{\infty}_{-\infty}dx\Biggl[ 
\left(\frac{d\phi}{dx}\right)^2+\left(\frac{d\chi}{dx}\right)^2+ 
W_{\phi}^2+W_{\chi}^2\Biggr] 
\ee 
It can be rewritten in the form 
\ben 
E&=&\frac{1}{2}\int^{\infty}_{-\infty}dx 
\Biggl[\left(\frac{d\phi}{dx}-W_{\phi}\right)^2+ 
\left(\frac{d\chi}{dx}-W_{\chi}\right)^2\Biggr]\nonumber 
\\ 
& &+\int^{\infty}_{-\infty}dx 
\left(W_{\phi}\frac{d\phi}{dx}+W_{\chi}\frac{d\chi}{dx}\right) 
\een 
We recall that 
\be 
\frac{dW}{dx}=W_{\phi}\frac{d\phi}{dx}+W_{\chi}\frac{d\chi}{dx} 
\ee 
to see that the energy gets to the bound $E_B$ when the field configurations 
obey the first-order equations (\ref{b1}) and (\ref{b2}). 
This is the Bogomol'nyi bound, and the corresponding 
solutions are BPS solutions. The BPS solutions are linearly 
or classically stable -- see Sec.~{\ref{sec:stability}}. 

The energy of BPS solutions can be also written as 
\ben 
\label{ee1} 
E_B&=&\int^{\infty}_{-\infty}dx\left(\left(\frac{d\phi}{dx}\right)^2+ 
\left(\frac{d\chi}{dx}\right)^2\right) 
\\ 
\label{ee2} 
&=&\int^{\infty}_{-\infty}dx\left(W^2_{\phi}+W^2_{\chi}\right) 
\een 
The above result shows that the gradient (g) and potential (p)
portions of the energy 
\ben 
& &{\rm g}=\int^{\infty}_{-\infty}dx\,{\rm g}(x)= 
\int^{\infty}_{-\infty}dx\,\frac{1}{2}\,
\Biggl[\left(\frac{d\phi}{dx}\right)^2+ 
\left(\frac{d\chi}{dx}\right)^2\Biggr] 
\\ 
& &{\rm p}=\int^{\infty}_{-\infty}dx\,{\rm p}(x)= 
\int^{\infty}_{-\infty}dx\,\frac{1}{2}\,\left(W^2_{\phi}+ 
W^2_{\chi}\right) 
\een 
contribute evenly to the total energy of the solution. 
Although this is shared by all the BPS states, it is not peculiar
to BPS states, since there are non-BPS states that also have energy
evenly distributed in its potential and gradient portions -- see below. 
 
In the case of two scalar fields, there are other possible arrangements 
to the energy (\ref{e}). We can, for instance, consider the possibility 
\ben 
E&=&\frac{1}{2}\int^{\infty}_{-\infty}dx 
\Biggl[\left(\frac{d\phi}{dx}-W_{\chi}\right)^2+ 
\left(\frac{d\chi}{dx}-W_{\phi}\right)^2\Biggr]\nonumber 
\\ 
& &+\int^{\infty}_{-\infty}dx 
\left(W_{\phi}\frac{d\chi}{dx}+W_{\chi}\frac{d\phi}{dx}\right) 
\een 
This expression is also minimized to obey Eqs.~(\ref{ee1}) and (\ref{ee2}) 
when one sets 
\ben 
\label{w1} 
\frac{d\phi}{dx}&=&W_{\chi} 
\\ 
\label{w2} 
\frac{d\chi}{dx}&=&W_{\phi} 
\een  
We can show that solutions to these equations (\ref{w1}) and (\ref{w2}) 
only solve the equations of motion when we impose on $W(\phi,\chi)$ 
the extra condition 
\be 
\label{ec1} 
W_{\phi\phi}=W_{\chi\chi} 
\ee 
Unfortunately, however, this condition (\ref{ec1}) factorizes
$W(\phi,\chi)$ into the sum $W_{+}(\phi_{+})+W_{-}(\phi_{-})$
when one rotates the $(\phi,\chi)$ plane 
to $(\phi_{+},\phi_{-})$, with 
\be 
\phi_{\pm}=\frac{1}{\sqrt{2}}(\chi\pm\phi) 
\ee 
as firstly shown in Ref.~{\cite{bbb99}}. The condition (\ref{ec1}) 
turns the system into two decoupled systems. 
 
We can consider another possibility to rewrite the energy (\ref{e}). 
We set 
\ben 
E&=&\frac{1}{2}\int^{\infty}_{-\infty}dx 
\Biggl[\left(\frac{d\phi}{dx}+W_{\chi}\right)^2+ 
\left(\frac{d\chi}{dx}-W_{\phi}\right)^2\Biggr]\nonumber 
\\ 
& &+\int^{\infty}_{-\infty}dx 
\left(W_{\phi}\frac{d\chi}{dx}-W_{\chi}\frac{d\phi}{dx}\right) 
\een 
This expression is also minimized to obey (\ref{ee1}) and (\ref{ee2}) 
when we write 
\ben 
\label{h1} 
\frac{d\phi}{dx}&=&-W_{\chi} 
\\ 
\label{h2} 
\frac{d\chi}{dx}&=&W_{\phi} 
\een 
These equations (\ref{h1}) and (\ref{h2}) solve the equations of motion 
when one imposes on $W(\phi,\chi)$ the extra condition 
\be 
\label{ec2} 
W_{\phi\phi}+W_{\chi\chi}=0 
\ee 
This result is interesting. It shows that for 
$W(\phi,\chi)$ harmonic, we can solve the two 
(second-order differential) equations of motion (\ref{x1}) and 
(\ref{x2}) with two sets of two first-order differential equations, 
the set of Bogomol'nyi equations (\ref{b1}) and (\ref{b2}), 
and another one, given by Eqs.~(\ref{h1}) and (\ref{h2}). 
 
We make good use of this result by imposing that the second set of 
first-order equations (\ref{h1}) and (\ref{h2}) are still Bogomol'nyi
equations, that appear via the presence of another function ${\widetilde W}$. 
This new function is defined by 
\ben 
\label{wbar1} 
& &{\widetilde W}_{\phi}=-W_{\chi} 
\\ 
\label{wbar2} 
& &{\widetilde W}_{\chi}=W_{\phi} 
\een 
Since $W$ and ${\widetilde W}$ are smooth functions, 
they are necessarily harmonic. We then get the result that potentials 
defined by harmonic functions can always be written in terms of two 
functions, ${W}$ and ${\widetilde W}$. We use this to introduce the complex 
superpotential 
\be 
{\cal W}=W+i\,{\widetilde W} 
\ee 
The superpotential ${\cal W}={\cal W}(\varphi)$ now depends on the complex
field $\varphi=\phi+i\,\chi$, and the harmonicity of both
$W$ and ${\widetilde W}$ appears via the Cauchy condition on ${\cal W}$.
This is the way we get from the investigations \cite{bds95,bsa96,brs96e}
to the recent work in Refs.~{\cite{gto99,cht99,saf99}}.

We guide ourselves toward the topological solutions by introducing a
topological current. This is the standard procedure, and we can introduce
for instance the topological current
\be 
\label{cb} 
J^{\alpha}_{BPS}=\varepsilon^{\alpha\beta}
\partial_{\beta}{\cal W}(\phi,\chi) 
\ee 
It is defined in terms of ${\cal W}(\phi,\chi)$, and the topological charge 
is directly identifyed with the energy of the topological solution that 
satisfies the pair of first-order equations (\ref{b1}) and (\ref{b2}). 
This definition identifies the topological charge with the energy 
of the BPS solution. It expresses the fact that the mass of a BPS 
state is completely determined by its topological charge.
This definition provides a single way to infer
stability of the BPS states. This is a nice way of
defining the topological current, but it is not complete, because it 
cannot see sectors of the non-BPS type. It is only effective for BPS 
sectors, and that is the reason for the label BPS assigned to it. 
However, since we need to identify every topological sector, we should 
use another topological current. We can consider, for instance 
\be 
\label{gc} 
J^{\alpha}=\varepsilon^{\alpha\beta}\partial_{\beta}{\phi\choose\chi} 
\ee 
It obeys $\partial_{\alpha}J^{\alpha}=0$, and it is also a vector in 
the $(\phi,\chi)$ plane. The topological current (\ref{gc}) is the current 
that we shall use in this work. We use this definition to obtain, 
for static configurations 
\be 
J^{t}_{\alpha}\,J^{\alpha}=\rho^t\,\rho=2\,{\rm g}(x) 
\ee 
where ${\rm g}(x)$ is the gradient energy density of the 
solution. This property can be used to infer stability of junctions, 
for configurations that present energy density evenly distributed
in their gradient and potential energy densities. This
is so because in this case $\rho^t\rho=2{\rm g}(x)={\rm g}(x)+{\rm p}(x)$
gives the full energy density, and may be used to show that the junction
process occurs exothermically. This result was already introduced in
Ref.~{\cite{bbr99a}}, and offers a way of showing how the process of junctions
of defects may occur exothermically in models defined outside the context of
supersymmetry -- see Sec.~{\ref{sec:junctions}}.

Up to here the calculations were done in $(1,1)$ space-time dimensions. 
However, we can easily immerse the system into $(2,1)$ or into $(3,1)$
space-time dimensions. In the first case, in $d=2$ the solutions can be
seen as domain ribbons, and the energy is now multiplied by the length
of the ribbons. In the second case, in $d=3$ we get domain walls, 
and the energy is to be multiplied by the area of the wall. The planar 
case is necessary for introducing junctions of domain ribbons [or of
domain walls if one immerses the system in the $(3,1)$ space-time] and
their corresponding planar networks. There are several distinct
possibilities, and in the sequel we specify some cases. We recall that
we single out topological sectors using pairs of vacuum states. For BPS
sectors the energy is obtained via the superpotential, and we shall also
use the notation $t_{ij}=|W(\phi_i,\chi_i)-W(\phi_j,\chi_j)|$,
identifying the BPS sectors with tensions or energies of the
topological defects.

\section{Some specific models}
\label{sec:models}

Our aim here is to introduce models of coupled scalar fields, 
to illustrate how one can find topological solutions 
of the BPS and non-BPS type. We investigate several models of coupled 
scalar fields, defined with smooth functions $W(\phi,\chi)$. 
We organize the subject in the four subsections that 
follow, where we examine four different models. 
 
\subsection{The first model} 
\label{sec:m1} 
 
We consider the superpotential
\be
\label{wm1} 
W^{(1)}(\phi,\chi)=-\lambda\phi+\frac{1}{3}\lambda\phi^3+\mu\phi\chi^2 
\ee
where $\lambda$ and $\mu$ are real parameters. This model was first
investigated in Ref.~{\cite{bds95}}, but here we bring 
new results. The equations of motion for static field configurations 
are given by 
\ben 
\label{m21} 
\frac{d^2\phi}{dx^2}&=& 2\lambda^2\phi(\phi^2-1) + 2\mu^2(2+ 
\frac{\lambda}{\mu}) \phi\chi^2 
\\ 
\label{m22} 
\frac{d^2\chi}{dx^2}&=& 2\lambda\mu(\phi^2-1)\chi+4\mu^2\phi^2\chi +  
2\mu^2\chi^3 
\een 
These equations are solved by configurations that also solve the pair 
of first-order differential equations 
\ben 
\label{b11} 
\frac{d\phi}{dx}&=&\lambda(\phi^2-1)+\mu\,\chi^2 
\\ 
\label{b12} 
\frac{d\chi}{dx}&=&2\,\mu\,\phi\,\chi 
\een 
Solutions that solve these first-order equations are BPS solutions. 
 
This model has four absolute minima when $\lambda/\mu>0$, 
and two minima when $\lambda/\mu<0$. For $\lambda/\mu>0$ the minima are 
$(\pm1,0)$ and $(0,\pm\sqrt{\lambda/\mu})$. Topological solutions are
solutions that connect adjacent minima. Thus, the asymptotic behavior
of the fields is governed by the set of minima 
of the potential. For $\lambda/\mu>0$ the system supports several 
topological sectors. Their energies or tensions are given by, 
for solutions that connect the minima $(-1,0)$ and $(1,0)$ 
\be 
\label{en1} 
t_{BPS}^{1}=\frac{4}{3}\,|\lambda| 
\ee 
For solutions that connect the minima $(\pm1,0)$ and 
$(0,\pm\sqrt{\lambda/\mu})$ we get 
\be 
\label{en2} 
{\bar t}_{BPS}^{\,1}=\frac{2}{3}\,|\lambda| 
\ee 
For the non-BPS solution that connects the minima $(0,\sqrt{\lambda/\mu})$ 
and $(0,-\sqrt{\lambda/\mu})$ we obtain 
\be 
\label{en3} 
t_{nBPS}^{\,1}=\frac{4}{3}\,|\lambda|\,\sqrt{\frac{\lambda}{\mu}} 
\ee 

There are BPS solutions in the first sector, which connects 
the two minima $(\pm1,0)$. They are
\ben 
\label{bs11} 
\phi(x)&=&-\tanh(\lambda\,x) 
\\ 
\label{bs12} 
\chi(x)&=&0 
\een 
and 
\ben 
\label{bs21} 
\phi(x)&=&-\tanh(2\,\mu\,x) 
\\ 
\label{bs22} 
\chi(x)&=&\pm\sqrt{\frac{\lambda}{\mu}-2\,}\,\,{\rm sech}(2\,\mu\,x) 
\een 
The first pair is of the one-component type, and is important for
investigating the presence of defects inside defects -- see for instance 
Refs.~{\cite{brs96,bba97,etb98,bbb99}}. The second pair is of 
the two-component type, and may play a role in applications to
condensed matter, as for instance in the case investigated in
Refs.~{\cite{brs96e,bve99}}.

The one-component and the two-component type of solutions
also appear in magnetic materials \cite{esc81}. There they are
known as Ising and Bloch walls, respectively.
The two-component Bloch wall has a peculiar behavior,
that can be used to understand specific features of interfaces between
magnetic domains. The issue here is that if one looks at these solutions
as vectors in the $(\phi,\chi)$ plane, for $x$ spanning the interval
$(-\infty,\infty)$ they describe a straight line and an elliptical arc
connection between the two minima $(\pm1,0)$, respectively, behaving
very much like linearly and elliptically polarized light. For this reason,
the Block walls can be seen as chiral interfaces between domains with
different magnetization, and this can be used to describe the wall
motion in magnetic materials \cite{cle90}. Furthermore, these
walls can be charged and we may introduce charge by making the
$\chi$ field complex, or by coupling fermions to the system.

For $\lambda/\mu<0$ we have only two vacuum states. They are $(\pm1,0)$ and 
now we have a single topological sector, with the same energy and topological 
charge obtained before, for the corresponding sector. 
We notice that the first pair of solutions (\ref{bs11}) and (\ref{bs12})
solves the first-order equations even for $\lambda/\mu<0$.
This one-component solution is still of the BPS type for $\lambda/\mu<0$,
but now no topological defect can be nested inside it. The reason is that
for $\lambda/\mu<0$ the model presents no spontaneous symmetry breaking
in the independent $\chi$ direction, and so it cannot support
topological defect in this $\chi$ direction anymore.

For $\lambda/\mu>0$, there are non-BPS solutions connecting 
$(0,\sqrt{\lambda/\mu})$ and $(0,-\sqrt{\lambda/\mu})$ by a straight 
line. To see this explicitly we consider $\phi=0$ in the specific system 
to get the equation of motion for the static field $\chi=\chi(x)$ 
\be 
\frac{d^2\chi}{dx^2}=2\mu^2\left(\chi^2-\frac{\lambda}{\mu}\right)\chi 
\ee 
The solutions are
\be 
\label{n1} 
\chi(x)=\pm\sqrt{\frac{\lambda}{\mu}\,}\tanh(\sqrt{\lambda\mu}\,x) 
\ee 
The corresponding non-BPS energy is
$(4/3)|\lambda|\,\sqrt{\lambda/\mu}$, as already presented
in Eq.~(\ref{en3}).

This model decouples for $\lambda=\mu$, as we have already
shown in Ref.~{\cite{bba97}}. In this case the symmetry $Z_2\times Z_2$
changes to the $Z_4$ one, but here a rotation by $\pi/4$ changes the
model into two single-field models. We also note that for $\lambda=-2\mu$,
the model decouples into two single-field systems, showing that the limit
$\mu\to-\lambda/2$ is also uninteresting, at least from the point of view
of models of coupled scalar field.

We expose other features of the model using $W^{(1)}$. We write 
\be
W^{(1)}_{\phi\phi}=2\lambda\phi,\,\,\,\,\,
W^{(1)}_{\chi\chi}=2\mu\phi,\,\,\,\,\, 
W^{(1)}_{\phi\chi}=2\mu\chi 
\ee 
We see that $W^{(1)}$ is harmonic when $\mu\to-\lambda$. This means that 
the model admits another pair of first-order equations. This pair follows 
from (\ref{h1}) and (\ref{h2}) 
\ben
\label{h11}
\frac{d\phi}{dx}&=&2\mu\phi\chi 
\\
\label{h21}
\frac{d\chi}{dx}&=&\lambda-\lambda\phi^2-\mu\chi^2 
\een 
These equations also furnish solutions to the equations of motion, 
for $\mu\to-\lambda$. As we have already seen, for $\mu\to-\lambda$
the model presents a single topological sector, already investigated.
But we notice that $W^{(1)}$ is also harmonic in the limit $\phi\to0$,
independently of the parameters $\lambda$ and $\mu$. We use this into
the above Eqs.~(\ref{h11}) and (\ref{h21}) to get $\phi=0$ and 
\be 
\frac{d\chi}{dx}=-\mu\left(\chi^2-\frac{\lambda}{\mu}\right) 
\ee 
We consider $\lambda\mu>0$. In this case the above equation
is easily solved to give 
\be 
\label{n11} 
\chi(x)=\sqrt{\frac{\lambda}{\mu}}\tanh(\sqrt{\lambda\mu}\,x) 
\ee 
The pair $\phi=0$ and $\chi$ in (\ref{n11}) is the pair 
of non-BPS solutions that we have already found. Here, however,
we found the non-BPS solution as an explicit solution of the new
first-order differential equations in systems of coupled real
scalar fields. Although this is a non-BPS solution, we have already
shown in Sec.~{\ref{sec:gmodels}} that its energy is composed of equal 
portions of gradient ({\rm g}) and potential ({\rm p}) contributions. 
In terms of densities we have, explicitly 
\be 
{\rm g}_1(x)={\rm p}_1(x)=\frac{1}{2}\lambda^2{\rm sech}^{4}
\left(\sqrt{\lambda\mu}\,x\right) 
\ee 
We can now understand this feature by realizing that although this pair
of solutions is non-BPS, in the reduced system where $\phi\to0$, the $\chi$ 
field solution is indeed a BPS solution.

\subsection{The second model} 
\label{sec:m2} 
 
We consider another model, defined by the superpotential 
\be 
\label{sm2} 
W^{(2)}(\phi,\chi)=\lambda(\phi^2-1)\chi 
\ee 
This model was also considered in Ref.~{\cite{csh97}}. The equations
of motion are
\be
\frac{\partial^2\phi}{\partial t^2}-
\frac{\partial^2\phi}{\partial x^2}+
2\lambda^2(\phi^2-1)\phi+4\lambda^2\phi\chi^2=0
\ee
and
\be
\frac{\partial^2\chi}{\partial t^2}- 
\frac{\partial^2\chi}{\partial x^2}+ 
4\lambda^2\phi^2\chi=0
\ee
For static fields the energy is minimized for configurations that obey the
first-order equations
\ben
\label{b31}
\frac{d\phi}{dx}&=&2\lambda\phi\chi 
\\
\label{b32}
\frac{d\chi}{dx}&=&\lambda(\phi^2-1)
\een

The model presents two vacuum states, given by $(\pm1,0)$. 
In contraposition with the former model, however, this new model 
presents no BPS sector, because the function $(\ref{sm2})$ 
obeys $W^{(2)}(\phi^2=1,\chi)=0$, and then gives $|\Delta W^{(2)}|=0$.
This means that the first-order equations (\ref{b31}) and (\ref{b32})
present no topological solutions connecting the two vacuum states.
On the other hand, the vacuum states $(\pm1,0)$ can be connected by 
topological but non-BPS solutions. The explicit form of the solutions 
is given by 
\ben 
\label{ns31} 
\phi(x)&=&\pm\tanh(\lambda\,x) 
\\ 
\label{ns32} 
\chi(x)&=&0 
\een 
The energy or tension of the solution is 
\be 
t_{nBPS}^{2}=(4/3)|\lambda| 
\ee 
This result reproduces the value presented by the similar one-component 
BPS solution (\ref{bs11}) and (\ref{bs12}) of the first model. 
 
We compare this model with the first one to see that it provides a nice way 
to understand how the BPS and non-BPS solutions behave. We notice that 
\be
W^{(2)}_{\phi\phi}=2\lambda\chi,\,\,\,\,\,
W^{(2)}_{\chi\chi}=0,\,\,\,\,\,
W^{(2)}_{\phi\chi}=2\lambda\phi 
\ee
There is only one way to make $W^{(2)}$ harmonic, which implies setting
$\chi\to0$. The new first-order equations correspond to the equations
(\ref{h1}) and (\ref{h2}) for $W^{(2)}$, in the limit $\chi\to0$. They give 
\be 
\frac{d\phi}{dx}=\lambda(\phi^2-1) 
\ee 
This is solved by 
\be 
\label{n2} 
\phi(x)=-\tanh(\lambda x) 
\ee 
The pair of solutions formed by $\chi=0$ and $\phi$ as in Eq.~(\ref{n2}) 
is the pair of non-BPS solutions that we have already found for this model. 
This is another example where one gets non-BPS solutions as solutions 
to first-order differential equations. This is perhaps the most natural
way of understanding why this non-BPS solution has energy evenly distributed
into its gradient and potential portions. In terms of energy densities
we have
\be 
{\rm g}_2(x)={\rm p}_2(x)=\frac{1}{2}\lambda^2{\rm sech}^{4}(\lambda\,x) 
\ee

\subsection{The third model}
\label{sec:m4}

Let us now consider another model, defined by the superpotential
\be
\label{ms}
W^{(3)}(\phi,\chi)=-\lambda\phi+\frac{1}{3}\mu\phi^3-\mu\phi\chi^2
\ee
This model has the first-order equations 
\ben 
\frac{d\phi}{dx}&=&-\lambda+\mu\phi^2-\mu\chi^2 
\\ 
\frac{d\chi}{dx}&=&-2\mu\phi\chi 
\een 
The model has only two minima, and for $\lambda/\mu>0$
they are $(\pm\sqrt{\lambda/\mu},0)$. They give
\be
W^{(3)}(\pm\sqrt{\lambda/\mu},0)=
\mp\frac{2}{3}\lambda\sqrt{\frac{\lambda}{\mu}}
\ee
and define a BPS sector. The explicit BPS solution is given by
\be
\label{bs+1} 
\phi(x)=-\sqrt{\frac{\lambda}{\mu}}\tanh\left(\sqrt{\lambda\mu}\,x\right), 
\ee
together with $\chi=0$. We deal with the first-order equations to get 
\be 
\chi\left(\phi^2-\frac{\lambda}{\mu}-\frac{1}{3}\chi^2\right)=0 
\ee 
This restriction constrains the orbits in the $(\phi,\chi)$ plane 
in such a way that no finite orbit can connect the two minima, 
unless in the case $\chi=0$, in which one reproduces the topological
solution just obtained. This model has a single topological sector,
and presents no family of topological solutions, like the family 
(\ref{bs21}) and (\ref{bs22}) obtained in the first model. 

We use $W^{(3)}$ to obtain 
\be 
W^{(3)}_{\phi\phi}=2\mu\phi,\,\,\,\,\,
W^{(3)}_{\chi\chi}=-2\mu\phi,\,\,\,\,\,
W^{(3)}_{\phi\chi}=-2\mu\chi 
\ee
We see that $W^{(3)}$ is harmonic. This means that 
we have another pair of first-order equations that come 
from (\ref{h1}) and (\ref{h2}). It is 
\ben 
\frac{d\phi}{dx}&=&2\mu\phi\chi 
\\ 
\frac{d\chi}{dx}&=&-\lambda-\mu\chi^2+\mu\phi^2 
\een 
We think of this pair as appearing from another function, 
${\widetilde W}^{(3)}$, as defined in the Eqs.~(\ref{wbar1})
and (\ref{wbar2}). We get 
\be 
\label{msbar} 
{\widetilde W}^{(3)}=-\lambda\chi-\frac{1}{3}\mu\chi^3+\mu\phi^2\chi 
\ee 
We notice that ${\widetilde W}^{(3)}$ vanishes along the line $\chi=0$ 
where the former sector was explicitly constructed.
 
We can use ${\widetilde W}^{(3)}$ to write ${\widetilde W}^{(3)}_{\phi
\phi}=2\mu\chi, {\widetilde W}^{(3)}_{\chi\chi}=-2\mu\chi,$ and
${\widetilde W}^{(3)}_{\phi\chi}=2\mu\phi$, showing that
${\widetilde W}^{(3)}$ is also harmonic, as expected. As we have
suggested in Sec.~{\ref{sec:gmodels}}, we use the complex superpotential
${\cal W}^{(3)}=W^{(3)}+i{\widetilde W}^{(3)}$ and the complex field
$\varphi=\phi+i\chi$. We change notation and write
\be
{\cal W}_{2}(\varphi)=\lambda\varphi-\frac{1}{3}\mu\varphi^3
\ee
as the superpotential of the third model. The new notation uses the subscript
$N$ in ${\cal W}_{N}$ to identify the complex superpotential that presents
the $Z_N$ symmetry -- see Sec.~{\ref{sec:junctions}} for further details.
We use ${\cal W}_{2}$ to get, in the vacuum states ${\bar\varphi}_k, k=1,2$
\ben
{\cal W}_2({\bar\varphi}_1)&=&-\frac{2}{3}\lambda\sqrt{\frac{\lambda}{\mu}}
\\
{\cal W}_2({\bar\varphi}_2)&=&\frac{2}{3}\lambda\sqrt{\frac{\lambda}{\mu}}
\een
or better
\be
{\cal W}_2({\bar\varphi}_k)=-\frac{2}{3}\lambda\sqrt{\frac{\lambda}{\mu}}\,
e^{2\pi\,i\,\frac{k-1}{2}}\,\,\,\,\,\,k=1,2
\ee
We use this result to obtain the tensions
\be
t^{(2)}_{jk}=|{\cal W}_2({\bar\varphi}_j)-{\cal W}_2({\bar\varphi}_k)|
\ee
In the present case there is a single topological sector, with tension
\be
t^{(2)}_1=\frac{4}{3}\,|\lambda|\,\sqrt{\frac{\lambda}{\mu}}
\ee
Here we use $t^{(N)}_n$ to identify tensions in models with complex
superpotentials having the $Z_N$ symmetry -- see Sec.~{\ref{sec:junctions}}
for more details.
\subsection{The fourth model} 
\label{sec:m5}
 
We consider another model, defined by the superpotential
\be 
\label{w4} 
W^{(4)}(\phi,\chi)={\bar\lambda}\phi+\bar{\mu}\chi-\frac{1}{4}\lambda\phi^4 
-\frac{1}{4}\lambda\chi^4+\frac{3}{2}\lambda\phi^2\chi^2 
\ee 
This model is similar to models already investigated in Ref.~{\cite{brs96e}}. 
The equations of motion are 
\ben 
\frac{d^2\phi}{dx^2}&=&6\, {\bar\mu}\lambda\,\phi\,\chi- 
3 {\bar\lambda}\lambda (\phi^2-\chi^2)-
6\lambda^2 \phi (\chi^2-3\phi^2)\chi^2\nonumber
\\
& &+3\lambda^2\phi(\phi^2-3\chi^2)(\phi^2-\chi^2)
\\ 
\frac{d^2\chi}{dx^2}&=&6\,{\bar\lambda}\lambda\,\phi\,\chi- 
3{\bar\mu}\lambda (\chi^2-\phi^2)-6\lambda^2\phi^2( \phi^2-3\chi^2)\chi
\nonumber
\\
& &
+3\lambda^2(\chi^2-3\phi^2)\,(\chi^2-\phi^2)\chi 
\een
These equations are solved by field configurations that obey the
first-order equations
\ben
\label{b41} 
\frac{d\phi}{dx}&=&{\bar\lambda}-\lambda\phi(\phi^2-3\chi^2) 
\\ 
\label{b42} 
\frac{d\chi}{dx}&=&{\bar\mu}-\lambda\chi(\chi^2-3\phi^2) 
\een 
 
The minima of the potential obey 
\ben 
& &\phi(\phi^2-3\chi^2)=\frac{\bar\lambda}{\lambda} 
\\ 
& &\chi(\chi^2-3\phi^2)=\frac{\bar\mu}{\lambda} 
\een 
Interesting cases are obtained with ${\bar\mu}=0$, and with 
${\bar\lambda}=0$. They give similar 
models, and we consider the case ${\bar\mu}=0$. We use
${\bar\phi}^3={\bar\lambda}/\lambda$, and the minima now obey 
\ben 
\phi(\phi^2-3\chi^2)&=&{\bar\phi}^3 
\\ 
\chi(\chi^2-3\phi^2)&=&0 
\een 
The minima are given by $(\phi_i,\chi_i), i=1,2,3$, where 
\ben\label{mw31} 
\phi_1={\bar\phi},& &\,\,\,\,\,\,\,\,\,\,\chi_1=0 
\\ 
\label{mw32} 
\phi_2=\phi_3=-\frac{1}{2}{\bar\phi},& &\,\,\,\,\,\, 
\chi_2=-\chi_3=\frac{\sqrt{3}}{2}{\bar\phi} 
\een 
They form a triangle in the $(\phi,\chi)$ plane. The ratio
${\bar\lambda}/\lambda$ specifies the vacuum states. For instance,
for ${\bar\lambda}=\lambda$ we get the minima at 
\be 
\left(1\, ,\,0\right),\,\,\,\,\,\, 
\left(\,-\,\frac{1}{2}\, ,\,\pm\,\frac{1}{2}\sqrt{3}\,\right) 
\ee 
 
The minima are equidistant from each other. The triangle they
form is equilateral, and the symmetry is the $Z_3$ symmetry.
The distance between the minima is $d=\sqrt{3}\,|{\bar\phi}|$,
and depend on the ratio ${\bar\lambda}/\lambda$. The function
$W^{(4)}$ is changed to, for ${\bar\mu}=0$,
\be 
\label{w3s} 
W^{(5)}={\bar\lambda}\phi-\frac{1}{4}\lambda\phi^4- 
\frac{1}{4}\lambda\chi^4+\frac{3}{2}\lambda\phi^2\chi^2 
\ee 
The equations of motion are solved by the first-order differential 
equations 
\ben 
\frac{d\phi}{dx}&=&\lambda[{\bar\phi}^3-\phi(\phi^2-3\chi^2)] 
\\ 
\frac{d\chi}{dx}&=&-\lambda\chi(\chi^2-3\phi^2) 
\een 
We use Eq.~(\ref{w3s}) to get 
\ben
& &W^{(5)}(\phi_1,\chi_1)=\frac{3}{4}\lambda{\bar\phi}^4
\\ 
& &W^{(5)}(\phi_2,\chi_2)=W^{(5)}(\phi_3,\chi_3)= 
-\frac{3}{8}\lambda{\bar\phi}^4 
\een

We notice that $W^{(4)}$ (and also $W^{(5)}$) is harmonic. This means that 
the potential of this model can be obtained by another superpotential,
given by 
\be 
\label{w41} 
{\widetilde W}^{(4)}(\phi,\chi)=-{\bar\mu}\phi+ 
{\bar\lambda}\chi-\lambda\phi^3\chi+\lambda\phi\chi^3 
\ee 
We set ${\bar\mu}=0$ to get 
\be 
\label{w4s} 
{\widetilde W}^{(5)}(\phi,\chi)={\bar\lambda}\chi- 
\lambda\phi^3\chi+\lambda\phi\chi^3 
\ee 
The corresponding first-order equations are 
\ben\label{be41} 
& &\frac{d\phi}{dx}=\lambda\chi(\chi^2-3\phi^2) 
\\ 
\label{be42} 
& &\frac{d\chi}{dx}=\lambda[{\bar\phi}^3-\phi(\phi^2-3\chi^2)] 
\een 
 
Evidently, the minima are $(\phi_i,\chi_i), i=1,2,3$, the same minima 
already obtained in Eqs.~(\ref{mw31}) and (\ref{mw32}). 
However, ${\widetilde W}^{(5)}$ behaves differently 
\ben 
& &{\widetilde W}^{(5)}(\phi_1,\chi_1)=0 
\\ 
& &{\widetilde W}^{(5)}(\phi_2,\chi_2)=
-\,{\widetilde W}^{(5)}(\phi_3,\chi_3)=
\,\frac{3}{8}\,\sqrt{3}\,\lambda\,{\bar\phi}^4 
\een 

This model leads to the model recently considered in
Refs.~{\cite{gto99,cht99}}. We proceed like in the form model
to get the complex superpotential
\be 
\label{wi}
{\cal W}_3(\varphi)={\bar\lambda}\,\varphi-
\frac{1}{4}\,{\lambda}\,\varphi^4 
\ee 
The field $\varphi$ is complex, $\varphi=\phi+i\chi$.
In Ref.~{\cite{gto99}} one considers the case ${\bar\lambda}=\lambda=1$.
In this case the potential is a particular case of the potential
introduced above. The superpotential can be seen as
${\cal W}_3=W^{(5)}+i\,{\widetilde W}^{(5)}$,
and this allows obtaining
\ben
{\cal W}_3({\bar\varphi}_1)&=&\frac{3}{4}\lambda{\bar\phi}^4
\\
{\cal W}_3({\bar\varphi}_2)&=&\frac{3}{4}\lambda{\bar\phi}^4\,
\left(-\frac{1}{2}+i\frac{1}{2}\sqrt{3}\right)
\\
{\cal W}_3({\bar\varphi}_3)&=&\frac{3}{4}\lambda{\bar\phi}^4\,
\left(-\frac{1}{2}-i\frac{1}{2}\sqrt{3}\right)
\een
or better
\be
{\cal W}_3({\bar\varphi}_k)=\frac{3}{4}\lambda{\bar\phi}^4\,e^{2\pi\,i\,
\frac{k-1}{3}},\,\,\,\,\,k=1,2,3
\ee
We use this result to write
\be
t^{(3)}_{jk}=|{\cal W}_3({\bar\varphi}_j)-{\cal W}_3({\bar\varphi}_k)|
\ee
The three tensions degenerate to the single value
\be
t^{(3)}_1=\frac{3}{4}\sqrt{3}|\lambda|{\bar\phi}^4
\ee

\section{Stability} 
\label{sec:stability} 
 
In this Sec.~{\ref{sec:stability}} we deal with issues concerning 
stability of the topological solutions in $(1,1)$ dimensions. 
We follow Ref.~{\cite{bbb99}}, recalling that the linear stability 
also shows how the bosonic fields behave in the background of the 
classical solutions. We examine linear stability for two coupled real 
scalar fields using $\phi(x,t)={\bar{\phi}}(x)+\eta(x,t)$ and 
$\chi(x,t)={\bar{\chi}}(x)+\xi(x,t)$. Here 
\ben 
\eta(x,t)&=&\sum_n\eta_n(x)\cos{(w_nt)}\\ 
\xi(x,t)&=&\sum_n\xi_n(x)\cos{(w_nt)} 
\een 
are the fluctuations around the static solutions ${\bar{\phi}}$ and 
${\bar{\chi}}$. We use the equations of motion to get the equation for 
stability  
\begin{eqnarray} 
\label{eig} 
\left( -\hat{1}\frac{{d}^2}{dx^2}+{U} \right)
{\eta_n\choose\xi_n}= {w}_n^2{\eta_n\choose\xi_n}
\end{eqnarray} 
For $w^2_n\geq 0$ $(w^2_n<0)$ we have stable (unstable) solutions. 
Here $\hat{1}$ is the $2\times2$ identity matrix and 
\begin{eqnarray} 
\label{M} 
{U}={ {{\bar{V}}_{\phi\phi}\,\,\,\,\,{\bar{V}}_{\phi\chi}}\choose 
{{\bar{V}}_{\chi\phi}\,\,\,\,\, {\bar{V}}_{\chi\chi}} }
\end{eqnarray} 
The bar over the potential means that after obtaining the derivatives, 
we should substitute the fields $\phi$ and $\chi$ by their classical 
static values $\bar{\phi}(x)$ and $\bar{\chi}(x)$. 
 
We consider the first model first. 
In this case the explicit calculation gives to the matrix $U^{(1)}$
the elements $U^{(1)}_{ij}=2\mu^2m_{ij}$, where
\ben 
& &m_{11}= 
{-\,\left(\frac{\lambda}{\mu}\right)^2+
3\left(\frac{\lambda}{\mu}\right)^2{\bar{\phi}}^2+ 
\left(\frac{\lambda}{\mu}+2\right)\,{\bar{\chi}}^2}
\\ 
& &m_{12}=m_{21}={2\,\left(\frac{\lambda}{\mu}+2\right)\,
{\bar{\phi}}\,{\bar{\chi}}}  
\\
& &m_{22}={-\,\frac{\lambda}{\mu}+\left(\frac{\lambda}{\mu}+2\right)\,
{\bar{\phi}}^2+3\,{\bar{\chi}}^2} 
\een
We use the first-order equations (\ref{b11}) and (\ref{b12}) to introduce 
the first-order operators 
\ben 
\label{S} 
S={\hat{1}}\frac{d}{dx}+2\mu{{{\frac{\lambda}{\mu}{\bar{\phi}}}
\,\,\,\,\,{\bar{\chi}}}\choose{{\bar{\chi}}\,\,\,\,\,{\bar{\phi}}}}
\een 
and 
\ben 
\label{Sdag} 
S^{\dag}=-{\hat{1}}\frac{d}{dx}+2\mu{ {\frac{\lambda}{\mu}{\bar{\phi}}}
\,\,\,\,\,{\bar{\chi}}\choose 
{{\bar{\chi}}\,\,\,\,\,{\bar{\phi}}} }
\een
These operators allow writing $H_1=S^{\dag}S$, where $H_1$ is the 
Hamiltonian corresponding to the first model  
\be 
H_1=-\hat{1}\frac{d^2}{dx^2}+U^{(1)} 
\ee 
This factorization $H_1=S^{\dag}S$ shows that $H_1$ is positive 
semi-definite, and so can have no negative eigenvalue. We then conclude 
that the BPS solutions are stable, since they solve the first-order 
equations we need to get the first-order operators $(\ref{S})$
and $(\ref{Sdag})$. 
 
In this first model, the one-component solutions can be solved explicitly. 
For the BPS pair (\ref{bs11}) and (\ref{bs12}) 
we get the following set of discrete eigenvalues: in the first direction, 
the $\phi$ direction 
\be 
w^{(1,1)}_0=0\,\,\,\,\,\,\,\,\,\,\,w^{(1,1)}_1=\sqrt{3\lambda^2} 
\ee 
and in the second direction, the $\chi$ direction 
\be 
w^{(1,2)}_0=0 
\ee 
For the pair of non-BPS solutions given by $\phi=0$ and $\chi$ in
Eq.~(\ref{n1}) we get
\ben
{\bar w}^{(1,1)}_n&=&\sqrt{\lambda\mu}\,\Biggl{\{}4-\Biggl[
\left(n+\frac{1}{2}\right)\nonumber
\\
& &-\sqrt{2\left(2+ 
\frac{\lambda}{\mu}\right)+\frac{1}{4}\,}\,\Biggr]^2\,\Biggr{\}}^{1/2}
\een
and
\be 
{\bar w}^{(1,2)}_0=0\,\,\,\,\,\,{\bar w}^{(1,2)}_1=\sqrt{3\lambda\mu} 
\ee 

Since $\lambda/\mu>0$, we find the restriction $0<\lambda/\mu\leq1$ 
on the ratio $\lambda/\mu$ in order to make the non-BPS solution stable. 
 
We notice that for $\mu=\lambda$ the eigenvalues ${\bar w}^{(1,1)}_n$ 
change to $0$ and $\sqrt{3\lambda^2}$, and degenerate to the values 
${\bar w}^{(1,2)}_n, n=0,1$, given above. This means that the scalar 
matter splits into two equal portions that 
present equal BPS states.

Let us now consider the second model. For the pair of non-BPS solutions 
given by Eqs.~(\ref{ns31}) and (\ref{ns32}) we get the following set of
eigenvalues for the bound states, in the first direction 
\be 
w^{(2,1)}_0=0,\,\,\,\,\,\,\,\,\,\,w^{(2,1)}_1=\sqrt{3\lambda^2} 
\ee 
In the second direction we have 
\ben 
w^{(2,2)}_0&=&\sqrt{[(\sqrt{17}-1)/2]\lambda^2} 
\\ 
w^{(2,2)}_1&=&\sqrt{[(3\sqrt{17}-5)/2]\lambda^2} 
\een 

Let us now consider the third model. We use the pair obtained with
(\ref{bs+1}) and $\chi=0$ to obtain for the bound states
the set of eigenvalues, in the first direction 
\be
w^{(3,1)}_0=0,\,\,\,\,\,\,\,\,\,\,w^{(3,1)}_1=\sqrt{3\lambda\mu} 
\ee
and, in the second direction
\be
w^{(3,2)}_0=\sqrt{2\lambda\mu} 
\ee

We cannot present explicit investigations concerning stability of
solutions of the fourth model, because we do not know any explicit
solution in this case. However, we consider stability of the solutions
of another model, introduced in Ref.~{\cite{bbr99a}}. This model is defined
by the superpotential
\ben 
W(\phi,\chi)&=&
\lambda\sqrt{\frac{4}{27}}\left(\chi+\sqrt{\frac{1}{3}}\right)^3
-\lambda\left(\chi+\sqrt{\frac{1}{3}}\right)^2\nonumber
\\
& &+\lambda\phi\left(\chi+\sqrt{\frac{1}{3}}\right)^2
-3\lambda\phi+\lambda\phi^3
\een
The system presents the three minima $(0,\pm\sqrt{4/3})$ and
$(\pm1,-\sqrt{1/3})$. They give three BPS sectors,
having energies $|\lambda|,\,3\,|\lambda|,$
and $4\,|\lambda|$. The highest energy is associated
to the sector connecting the minima $(\pm1,-\sqrt{1/3})$.
This is the only sector where we can find explicit solutions.
They are given by 
\ben 
& &\phi(x)=-\tanh(3\lambda\,x) 
\\ 
& &\chi(x)=-\sqrt{1/3\,} 
\een 
This pair of solutions represents a straight line orbit
in the $(\phi,\chi)$ plane, connecting the two vacua $(\pm1,-\sqrt{1/3})$. 
We consider stability of the above solutions. Here the fluctuations
decouple and we obtain
\ben
U_{11}&=&9\lambda^2[4-6\,{\rm sech}^2(3\lambda x)]
\\
U_{22}&=&9\lambda^2\left(\frac{8}{9}\right)
\left[1+\tanh(3\lambda x)-\frac{5}{4}{\rm sech}^2(3\lambda x)\right]
\een
The first potential $U_{11}$ is the modified P\"osch-Teller potential.
It presents two bound states, with energies
\be
w_0=0,\,\,\,\,\,\,\,\,\,\,w_1=3\sqrt{3\lambda^2}
\ee
The second potential $U_{22}$ is not reflectionless. It supports
no bound state, and the continuum starts at zero energy.

\section{Further considerations}
\label{sec:further}

The models investigated above present similar BPS and non-BPS solutions.
They provide concrete ways of investigating how the scalar fields behave
in the background of BPS and non-BPS solutions. We explore some possibilities
in the next Sec.~{\ref{sec:bag}}. In Sec.~{\ref{sec:junctions}} we
investigate intersection of defects.

\subsection{Entrapment of the other field} 
\label{sec:bag} 
 
In the first model, we can further understand the importance of the 
topological solutions by investigating the potential 
$V_1(\phi,\chi)$. It is given by
\ben
V_1(\phi,\chi)&=& \frac{1}{2}\lambda^2(\phi^2 - 1)^2+ 
\lambda\mu(\phi^2-1)\chi^2\nonumber
\\
& &+\frac{1}{2}\mu^2\chi^4+2\mu^2\phi^2\chi^2 
\een
This model presents the one-component BPS solution (\ref{bs11}) and 
(\ref{bs12}). We have $\phi^2=1$ outside the BPS defect, and $\phi=0$ 
at $x=0$, at the center of the defect. We then use $V_1(\phi,\chi)$
to get 
\ben 
\label{potchi} 
& &V_1(\phi\to0,\chi)=\frac{1}{2}\mu^2\left(\chi^2-
\frac{\lambda}{\mu}\right)^2 
\\ 
\label{potchi1} 
& &V_1(\phi^2\to1,\chi)=2\mu^2\chi^2+\frac{1}{2}\mu^2\chi^4 
\een
For $\lambda/\mu>0$ the potential (\ref{potchi}) shows the presence 
of spontaneous symmetry breaking in the independent $\chi$ direction 
in the $(\phi,\chi)$ plane. In this case the $\chi$ field presents 
squared masses given by 
\ben 
m^2_{\chi}(in)&=&4\lambda\mu 
\\ 
m^2_{\chi}(out)&=&4\mu^2 
\een 
where $in$ and $out$ identify the region inside and outside the defect. 
We see that the ratio of masses is 
\be 
\label{ratio} 
\frac{m^2_{\chi}(in)}{m^2_{\chi}(out)}=\frac{\lambda}{\mu} 
\ee 
This ratio identifies the region $0<\lambda/\mu<1$ in the space of 
parameters where the field $\phi$ entraps the other field. 
For a given $\lambda$, we see that the parameter $\mu$
controls the ratio of squared masses. Thus, the efficiency of the 
entrapment depends on $\mu$. 
 
This model can be used as a bag model \cite{cjj74}, 
similar to the models in Refs.~{\cite{rwe75,mon76}}. 
Here we can further enlarge the model by changing dimensions from 
$(1,1)$ to $(3,1)$ and making the $\chi$ field complex, 
going from the $Z_2\times Z_2$ symmetry to the $Z_2\times U(1)$. However, 
since inside the defect formed by $\phi(x)$, 
the $\chi$ field still engenders spontaneous symmetry breaking for
$\lambda\mu>0$, we can only circumvent the presence of Goldstone
bosons when one gauge the 
related $U(1)$ symmetry, making it local. The model that emerges 
presents the symmetry $Z_2\times U_l(1)$. Since we started in $(3,1)$ 
dimensions, inside the wall the theory is effectively $(2,1)$ 
dimensional. This is an alternate mechanism for dimensional 
reduction, and inside the wall the effective theory may very naturally 
develop the Callan-Harvey effect \cite{cha85}, if one appropriately
couples fermions to the second field, $\chi$. This scenario seems to
appear interesting, and we shall further explore this possibility
in a separate report. 
 
The first model is simpler for $\lambda\mu<0$. In this case, spontaneous 
symmetry breaking no longer appear inside the wall formed by the $\phi$ 
field. However, the wall is still of the BPS type. The potential 
$V_1(\phi,\chi)$ changes in a way such that the ratio (\ref{ratio}) becomes 
\be 
\frac{m^2_{\chi}(in)}{m^2_{\chi}(out)}=-\frac{\lambda}{2\mu} 
\ee 
This result shows that the topological defect generated by $\phi$ 
can still entrap $\chi$ mesons for $0>\lambda/\mu>-2$, acting 
like a bag for the elementary excitations of the field $\chi$. 
If we change the symmetry $Z_2\times Z_2$ to $Z_2\times U(1)$ by making 
the $\chi$ field complex, in this case we do not need to make the $U(1)$
symmetry local, and we can still explore the presence of a global
symmetry within the wall. Work in this direction was already presented in
Ref.~{\cite{pet96}}, in a model inspired in the work of Ref.~{\cite{wit85}},
with a potential that differs from the one here proposed. We believe that
our model may give further insight into the issues that arise in this
scenario. Further applications can be done, as for instance in
Ref.~{\cite{bar97}}, where one explores the presence of diffuse
domain walls at intersecting flat directions within the cosmological
scenario. Also, we can follow Ref.~{\cite{gbl98}} to investigate issues
concerning hybrid inflation, envolving features that requires at least
two real scalar fields, the inflaton field and another one, which is in
general used to break the symmetry.

The second model is defined with the superpotential (\ref{sm2}). It gives
the potential
\be
V_2(\phi,\chi)=\frac{1}{2}\lambda^2(\phi^2-1)^2+2\lambda^2\phi^2\chi^2 
\ee
Here we have
\ben
V_2(\phi=0,\chi)&=&\frac{1}{2}\lambda^2
\\
V_2(\phi^2=1,\chi)&=&2\lambda^2\chi^2
\een 
Thus the squared masses associated to the $\chi$ field are 
\ben 
m^2_{\chi}(in)&=&0 
\\ 
m^2_{\chi}(out)&=&4\lambda^2 
\een 
In this case the non-BPS defect always entraps the field $\chi$, acting 
like a bag for the elementary $\chi$ excitations. Here, however, 
the $\chi$ meson is massless inside the bag. This shows that the entrapment 
is more efficient in this model, although the basic solution, the bag, 
is of the non-BPS type. 

We can investigate the third model similarly. It is defined by the
superpotential (\ref{ms}) and we have
\ben 
\label{p2} 
V_3(\phi,\chi)&=&\frac{1}{2}\mu^2\left(\phi^2-
\frac{\lambda}{\mu}\right)^2-\mu^2\left(\phi^2-
\frac{\lambda}{\mu}\right)\chi^2\nonumber
\\
& &+\frac{1}{2}\mu^2\chi^4+2\mu^2\phi^2\chi^2 
\een 
We can write, for $\lambda/\mu>0$, 
\ben 
& &V_3(0,\chi)=\frac{1}{2}\mu^2\left(\chi^2+ 
\frac{\lambda}{\mu}\right)^2 
\\ 
& &V_3(\pm\sqrt{\lambda/\mu},\chi)=2\lambda\mu\chi^2+ 
\frac{1}{2}\mu^2\chi^4 
\een 
No spontaneous symmetry breaking appears inside the $\phi$-defect
anymore, and so the model cannot support defect inside defect. However, 
the squared masses of the elementary $\chi$ mesons are 
\ben
{\tilde m}^2_{\chi}(in)&=&2\lambda\mu
\\
{\tilde m}^2_{\chi}(out)&=&4\lambda\mu
\een
We get the ratio
\be
\frac{{\tilde m}^2_{\chi(in)}}{{\tilde m}^2_{\chi(out)}}=\frac{1}{2}
\ee
This ratio does not depend on the parameters $\lambda$ and $\mu$, 
but the topological defect may also entrap $\chi$ mesons.
This is another model,
in which the topological state may simulate a bag to entrap the elementary 
excitations of the other field. In this case, however, the efficiency 
of the entrapment can no longer be controlled by the parameters that 
defines the model.

The second and third models may also be immersed in the $(3,1)$ 
dimensional space-time. This immersion provide alternative ways 
to investigate the behavior of the current generated by making the
second field complex, as commented above. They give similar scenarios
for charged walls, and may present distinct physical contents.

Other applications include the diverse issues recently
investigated in Refs.~{\cite{bba97,mor98,etb98,bbb99,mba96,mor99}}.
We can, for instance, enlarge the models in order to break the symmetry
and/or the supersymmetry. This possibility may give rise to Fermi
balls \cite{mba96}, Fermi disks \cite{etb98}, and other charged
objects \cite{mor99}, which depend on the particular space-time
dimension one starts with, and on the parameters that define
the terms added to allow the symmetry breaking.
We can also follow Refs.~{\cite{bba97,mor98,bbb99}} to extend the
scenario of nested defects to the case of nested network of defects.
This picture follows as a direct generalization of the result of
Ref.~{\cite{bbb99}}, which shows explicitly that defects may nest
defects if the model engenders the $Z_2\times Z_2$ symmetry in
systems of two real scalar fields. As we are going to see below,
to have intersection of defects we must necessarily enlarge the $Z_2$
symmetry to at least the $Z_3$ one. For this reason, we can think of a
model that presents the $Z_2\times Z_3$ symmetry, composed of three real
scalar fields, one presenting the $Z_2$ symmetry, and the other two
controlling the $Z_3$ symmetry. In $(3,1)$ space-time dimensions
the first field engenders the $Z_2$ symmetry, and we can certainly
get to the case where it may give rise to a domain wall, which may
nest the two other fields in its interior. Within this scenario,
if the $Z_3$ symmetry related to the second and third fields is still
effective we may nest a planar network of defects inside domain walls.
We take advantage of the model investigated in Ref.~{\cite{bbr99a}}
to present a model that seems to work correctly.
This model is defined by
\ben
V(\psi,\phi,\chi)&=&\frac{2}{3}\lambda^2\left(\psi^2-\frac{9}{4}\right)^2
-\lambda^2\psi^2\,(\phi^2+\chi^2)
\nonumber\\
& &+\lambda^2(\phi^2+\chi^2)^2-\lambda^2\phi(\phi^2-3\chi^2)
\een
It behaves standardly in $(3,1)$ space-time dimensions and is such that
$V(\psi)=V(\psi,0,0)$ obeys
\be
V(\psi)=\frac{2}{3}\lambda^2\left(\psi^2-\frac{9}{4}\right)^2
\ee
We consider that the field $\psi=\psi(x)$ depends only on $x$. Thus,
it supports BPS defects with the explicit form
\be
\psi(x)=-\frac{3}{2}\,\tanh\left(\sqrt{3}\,\lambda\,x\right)
\ee
The corresponding energy density is $3\sqrt{3\,\lambda^2}$.
We also notice that the potentials $V_{in}(\phi,\chi)=V(0,\phi,\chi)$
and $V_{out}(\phi,\chi)=V(\pm3/2,\phi,\chi)$ are such that
\be
V_{in}(\phi,\chi)=\frac{27}{8}\lambda^2+\lambda^2(\phi^2+\chi^2)^2
-\lambda^2\phi(\phi^2-3\chi^2)
\ee
and
\ben
V_{out}(\phi,\chi)&=&\lambda^2(\phi^2+\chi^2)^2-
\lambda^2\phi(\phi^2-3\chi^2)\nonumber\\
& &-\frac{9}{4}\lambda^2(\phi^2+\chi^2)
\een
We see that the $Z_3$ symmetry of the pair of fields $(\phi,\chi)$ is
preserved both inside and outside the domain wall generated by the first
field, $\psi$. In the region inside the domain wall, $V_{in}(\phi,\chi)$
has minima at the points
\be
\left(\frac{3}{4},0\right),\,\,\,\,\,
\left(-\frac{3}{8},\pm\frac{3}{8}\sqrt{3}\right)
\ee
In this region inside the domain wall the squared masses corresponding to
the two mesons are
\ben
m^2_{\phi}(in)&=&\frac{9}{4}\lambda^2
\\
m^2_{\chi}(in)&=&\frac{27}{4}\lambda^2
\een
In the region outside the domain wall, $V_{out}(\phi,\chi)$
has the minima \cite{bbr99a}
\be
\left(\frac{3}{2},0\right)\,\,\,\,\,\,\,
\left(-\frac{3}{4},\pm\frac{3}{4}\sqrt{3}\right)
\ee
In this case the squared masses are
$m^2_{\phi}(out)=m^2_{\chi}(out)=(27/2)\,\lambda^2$.
These results show that the field $\psi$ engenders an efficient mechanism
for the entrapment of the pair of fields $(\phi,\chi)$ in this model. We
shall further explore this and other related issues in another work.

\subsection{Intersection of defects}
\label{sec:junctions}

The process of intersection of defects was already explored in
Ref.~{\cite{ato91}}, in the case of supersymmetric field theories.
There, the basic idea was to show that the intersection of two extended
objects is energetically admissible when it can be seen as a reaction
that occurs exothermically. This may be translated into the expression 
$t_{ik}<t_{ij}+t_{jk}$, where $t_{ik},\,\, t_{ij}$, and $t_{jk}$ are the
energies or tensions associated to the fusion of the two defects, and to
the individuals defects, respectively. The crucial point here is that in
supersymmetric theories the topological charge ${\cal T}$ of the defect
appears as a central charge, and for BPS states one gets $t=|{\cal T}|$.
If one uses $t_{ij}=|{\cal T}_{ij}|, t_{jk}=|{\cal T}_{jk}|$,
and $t_{ik}=|{\cal T}_{ij}+{\cal T}_{jk}|$ the triangle
inequality $t_{ik}\leq t_{ij}+t_{jk}$ follows naturally.
This triangle inequality is strictly valid in supersymmetric
systems described by complex superpotentials, whose values at
the vacuum states define points in the complex plane, not aligned
into a straight-line segment.

The fourth model is the supersymmetric model that opens the possibility 
for the presence of stable intersection of extended objects, giving rise
to a triple junction of BPS states, as recently shown in Ref.~{\cite{gto99}}
-- see also Refs.~{\cite{cht99,saf99}}. This model was considered
in Sec~{\ref{sec:m5}}, and the basic model is given by
$\varphi-(1/4)\,\varphi^4$. Here, however, we consider the case
\be
{\cal W}_N={\bar\lambda}\,\varphi-\frac{1}{N+1}\,\lambda\,\varphi^{(N+1)}
\ee
with $\lambda$ and ${\bar\lambda}$ real parameters. This model presents
the $Z_N$ symmetry. The vacuum states are singular points of the
superpotential. They obey $\varphi^N={\bar\lambda}/\lambda$, and the $N$
solutions are, using ${\bar\lambda}=a^{N}\lambda$ with $a$ real and positive,
\be
{\bar\varphi}_k=a\,e^{2\pi i\left[ (k-1)/N\right]},\,\,\,\,\, k=1,2,...,N.
\ee
The topological sectors connect pairs of adjacent vacua, and in general the
number of BPS sectors is $N(N-1)/2$. Their energies or tensions are given in
terms of the values of the superpotential in the singular points, that is
\be
t^{(N)}_{ij}=\left|{\cal W}_N({\bar\varphi}_i)-
{\cal W}_N({\bar\varphi}_j)\right|
\ee
The tensions can be classified according to the integer $n$,
$n=1,2,...,\leq[N/2]$, where $[N/2]$ stands for the greatest
integer not greater than $N/2$ itself. These tensions have the
explicit values
\be
t^{(N)}_n=\frac{Na^{(N+1)}}{N+1}\,|\lambda|
\sqrt{2\left[1-\cos\left(2\pi\frac{n}{N}\right)\right]}
\ee
The integer $n$ identifies topological sectors described via
the connection between a given vacuum, and its $n-th$ neighbor:
$n=1$ for connections between first neighbors, $n=2$ for second
neighbors, and so forth. We can check that these tensions are ordered
in the form $t^{(N)}_k<t^{(N)}_{k+1},$ for $k=1,...,<[N/2]$, and $N\geq4$.
This order also shows that the direct connection between two
vacuum states is always less energetic than any other possible
connection. For instance, $t^{(N)}_{2}<2\,t^{(N)}_1,\,N\geq4$,
and $t^{(N)}_3<t^{(N)}_2+t^{(N)}_1<3\,t^{(N)}_1,\,N\geq6$, and so forth.

More recently, in Ref.~{\cite{bbr99a}} we considered another model.
The model is defined by the potential
\ben 
\label{p+} 
V(\phi,\chi)&=&\lambda^2\phi^2\left(\phi^2-
\frac{9}{4}\right)+\lambda^2\chi^2\left(\chi^2-
\frac{9}{4}\right)\nonumber 
\\ 
& &+\,2\,\lambda^2\,\phi^2\,\chi^2-\lambda^2\phi\,(\phi^2-
3\,\chi^2)+\frac{27}{8}\lambda^2 
\een
It presents the $Z_3$ symmetry, but it is described by a potential
that cannot be written in terms of some superpotential.
In this case the key issue relies on finding a way out
of supersymmetry to show that the fusion of defects still
occurs exothermically. In the supersymmetric
case one can use interesting properties to make the
reasoning naturally simple. In the non-supersymmetric case, however,
we have found non-BPS solutions that presents the interesting property
of having gradient and potential portions of the energy evenly distributed
to compose the total energy of the defect. As we have shown,
such a property is also shared by all the BPS solutions,
and so we can use it to present a new reasoning, that
works within and beyond the context of supersymmetry. The reasoning follows
from the topological current presented in Eq.~(\ref{gc}). Here we have
\be
\label{rhoz3}
\rho(x)={ {\phi'(x)}\choose{\chi\, '(x)} }
\ee
Thus $\rho^t\rho=2{\rm g}(x)$, and for static solutions that presents equal
gradient and potential portions we get
$\rho^t\rho={\rm g}(x)+{\rm p}(x)=\varepsilon(x)$, or better 
\be
\label{rhoint}
t_{ij}=\int^{\infty}_{-\infty}dx\,\rho^t_{ij}\,\rho_{ij}
\ee
We consider fusion of two defects in the model investigated in
Ref.~{\cite{bbr99a}}. The topological defects there obtained
obey ${\rm g}(x)={\rm p}(x)$. Thus, we can use
\ben
\label{inter}
(\rho_{ij}+\rho_{jk})^t(\rho_{ij}+\rho_{jk})&=&
\rho_{ij}^t\rho_{ij}+\rho_{jk}^t\rho_{jk}\nonumber
\\
& &+\rho_{ij}^t\rho_{jk}+\rho_{jk}^t\rho_{ij}
\een
Also, we take advantage of the symmetry of the model to see that if
the pair $(\phi,\chi)$ represents a solution in a given sector,
all the other solutions can be obtained by rotating the solution
according to the $Z_3$ symmetry. This means that we can write
\be
\label{rhoz31}
{ {\phi'_{ij}(x)}\choose{\chi'_{ij}(x)} }
={{\,\,\,\,\,\cos(\alpha_{\,ij})\,\,\,\,\,\sin(\alpha_{\,ij})}\choose
{-\sin(\alpha_{\,ij})\,\,\,\,\,\cos(\alpha_{\,ij})}}\,
{ {\phi'(x)}\choose{\chi'(x)} }
\ee
Here $\alpha_{\,ij}$ is the angle between the given solution $(\phi,\chi)$
and the solution $(\phi_{ij},\chi_{ij})$. In the $Z_3$ model under
consideration $\alpha_{ij}$ can only be $2\pi/3$ and $4\pi/3$.
We now use (\ref{rhoint}), (\ref{inter}), and (\ref{rhoz31}) to obtain
$t_{ik}<t_{ij}+t_{jk}$. This result shows that the three-junction in the
model under consideration occurs exothermically.

\section{Networks of defects} 
\label{sec:network} 
 
The recent Refs.~{\cite{gto99,cht99,saf99}} have introduced investigations 
concerning defect junctions and networks in supersymmetric systems,
and now we turn our attention to this possibility. We consider 
junctions in the plane, which may give rise to planar networks of
defects. To implement this possibility we work in the $(2,1)$ space-time, 
in the plane $(x,y)$. We identify the axes $(x,y)$ with the axes
$(\phi,\chi)$ of the space of configurations. We illustrate
this situation by considering the model defined by the potential
in Eq.~(\ref{p+}). The topological solutions in the planar
case are \cite{bbr99a}
\ben
\phi^{\pm}_{(2,3)}&=&-\frac{3}{4}
\\
\chi^{\pm}_{(2,3)}&=&\pm\frac{3}{4}\sqrt{3}
\tanh\left(\sqrt{\frac{27}{8}}\,\lambda\,y\right)
\een
and
\ben
\phi^{\pm}_{(1,2)}&=&\frac{3}{8}\Biggl[1\mp 3\tanh\left(
\frac{1}{2}\sqrt{\frac{27}{8}}\,\lambda\,(y+\sqrt{3}x)\right)\Biggr] 
\\
\chi^{\pm}_{(1,2)}&=&\frac{3}{8}\sqrt{3}\Biggl[1\pm\tanh\left(
\frac{1}{2}\sqrt{\frac{27}{8}}\,\lambda\,(y+\sqrt{3}x)\right)\Biggr] 
\een 
and 
\ben 
\phi^{\pm}_{(1,3)}&=&\frac{3}{8}\Biggl[1\pm 3\tanh\left(
\frac{1}{2}\sqrt{\frac{27}{8}}\,\lambda\,(y-\sqrt{3}x)\right)\Biggr] 
\\ 
\chi^{\pm}_{(1,3)}&=&-\frac{3}{8}\sqrt{3}\Biggl[1\mp\tanh\left(
\frac{1}{2}\sqrt{\frac{27}{8}}\,\lambda\,(y-\sqrt{3}x)\right)\Biggr]
\een
They may be used to represent the three-junction in the thin wall
approximation.

Let us consider the first model, in the case $\lambda/\mu>0$.
The minima are at $v_1=(1,0)$, $v_2=(0,\sqrt{\lambda/\mu})$, $v_3=(-1,0)$, 
and $v_4=(0,-\sqrt{\lambda/\mu})$. The energies or tensions of the BPS and
non-BPS sectors are
$t^1_{BPS}=2{\bar t\,}^1_{BPS}=\sqrt{(\mu/\lambda)}t^1_{nBPS}=(4/3)|\lambda|$. 
For a given $\lambda$, the value of $\mu$ allows obtaining three distinct
situations. The first concerns the case $\mu>\lambda$. In this case the 
minima give rise to an array of defects of the form shown in FIG.~1, in
the thin wall approximation.

\begin{figure} 
\centerline{\psfig{figure=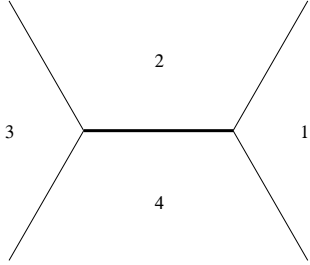,height=3.5cm}}
\vspace{1.0cm} 
\caption{A possible array of defects, as suggested by the first model
of Sec.~{\ref{sec:m1}} in the case of $\lambda/\mu<1$. The vacuum states
are represented by $v_1=1, v_2=2,v_3=3$, and $v_4=4$, and the thinner and
thicker lines stand for BPS and non-BPS defects, respectively.}
\end{figure}

The second case concerns $\mu=\lambda$, and the third one $\mu<\lambda$. 
In the thin wall approximation they are shown in FIG.~2, and in FIG.~3,
respectively. The second case seems to give an intersection of four 
half-defects, but this is not true since for $\mu=\lambda$ the two fields 
decouple. Thus, the apparent intersection of four half-defects represents
in fact the uninteresting case of two non-intersecting defects.

The first and third cases seem to be dual to each other,
since they are linked by the interchange 
$\lambda\leftrightarrow\mu$. This duality is very much like the 
$s\leftrightarrow t$ duality that the diagrams depicted in FIG.~1
and in FIG.~3 remember. This scenario excludes the self-dual case, 
where $\mu=\lambda$.

The basic diagrams of FIG.~1 and FIG.~3 can be used to generate planar 
networks of defects. We can generate a network in the two following ways. 
The s-network, which appears for $\lambda/\mu<1$, and the t-network,
for $\lambda/\mu>1$. They are depicted in FIG.~4 and in FIG.~5, in the
case of regular hexagonal networks in the thin wall approximation.
In the first case we impose that the triangle with vertices $v_1,v_2,v_4$
is equilateral. This gives $\lambda/\mu=1/3$. In the second case
we impose that the triangle with vertices $v_1,v_2,v_3$ is equilateral.
It gives $\lambda/\mu=3$, the result we should expect from duality.
We notice that the first (second) imposition subdivides the distance
between the horizontal (vertical) vacua into three equal pieces,
which is the condition for the formation of a network of regular hexagons.

\begin{figure} 
\centerline{\psfig{figure=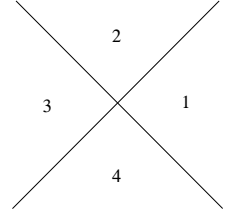,height=2.8cm}}
\vspace{1.0cm} 
\caption{Another array of defects, suggested by the first model of
Sec.~{\ref{sec:m1}} in the case of $\lambda/\mu=1$.} 
\end{figure}

\begin{figure} 
\centerline{\psfig{figure=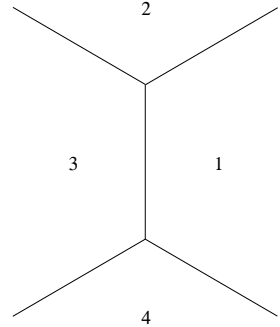,height=4.3cm}}
\vspace{1.0cm} 
\caption{Another possible array of defects, as suggested by the first model
of Sec.~{\ref{sec:m1}} in the case of $\lambda/\mu>1$.} 
\end{figure}

For $\mu=3\lambda$, we have two sides
of the hexagon formed by non-BPS states, with tension $\sqrt{(1/3)}t^1_{BPS}$.
The other four sides are formed by BPS states, with tension $(1/2)t^1_{BPS}$.
For $\mu=(1/3)\lambda$, two sides of the hexagon are formed by
BPS states, with tension $t^1_{BPS}$, and the other four sides are also
formed by BPS states, with tension $(1/2)\,t^1_{BPS}$.

We think of a planar network of defects. We notice that a t-network means
$\lambda/\mu>1$, and this allows hexagonal networks with two
distinct BPS states, having two different tensions, $t^1_{BPS}$
and ${\bar t\,}^1_{BPS}=(1/2)t^1_{BPS}$, showing that each three-junction
of this t-network is at the threshold of stability. For $\lambda/\mu<1$,
in the case of s-networks, however, the situation is different since now
we must have two non-BPS states to build the hexagonal cell. Since the
ratio $\lambda/\mu<1$ controls the energy of the non-BPS state, every
three-junction may obey the inequality $t_{ij}<t_{jk}+t_{ki}$.

\begin{figure} 
\centerline{\psfig{figure=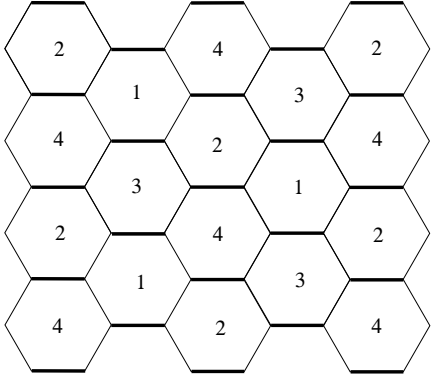,height=5.0cm}} 
\vspace{1.0cm} 
\caption{The s-network, depicted as a regular hexagonal network
of defects in the case of $\lambda/\mu=1/3$. The thicker lines
represent non-BPS states.}
\end{figure}

\begin{figure} 
\centerline{\psfig{figure=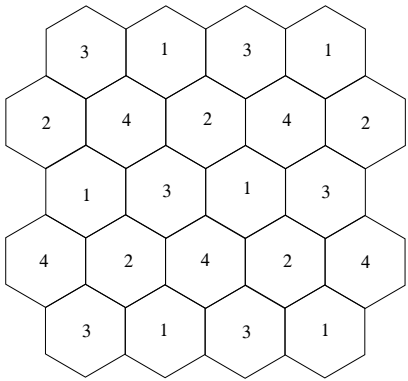,height=5.0cm}} 
\vspace{1.0cm} 
\caption{The t-network, depicted as a regular hexagonal network of
defects in the case of $\lambda/\mu=3$.}
\end{figure}

We can consider the case where $t^1_{nBPS}={\bar t\,}^1_{BPS}$.
This gives $\mu=4\,\lambda$ and the tensions degenerate to the single
value $(2/3)|\lambda|$. The hexagonal cell is no longer regular.
The angle between the two BPS states changes from $\alpha=120$ to
$\alpha=2\,\theta$, where $\theta$ is slightly lesser than $63.5$
degrees. We notice that we cannot make the tensions and angles degenerate
simultaneously, and this indicates instability of the suggested networks.

Effects that contribute to stabilize the network may appear when we go
beyond the classical level and introduce the quantum corrections, since the
quantum contributions may modify the classical scenario. For instance,
in the case $\lambda/\mu>1$, in a t-network all the defects are BPS defects,
and their energies or tensions are protected against receiving quantum
corrections. On the other hand, for $\lambda/\mu<1$, in the case of a
s-network there are non-BPS states whose tensions are not protected and
may receive quantum corrections, changing the classical value $t^1_{nBPS}$,
contributing to stabilize the network.

The first model does not allow the appearence of a network 
of planar squares, although squares can also fill the plane.
In our model this would require the $Z_4$ symmetry, which is 
not effective in that model.

The fourth model leads to the model considered in
Refs.~{\cite{gto99,cht99}}. In this case the three-junction
is formed by BPS defects, and we can form an hexagonal array.
Here, the three-junction is stable, because each leg carries the
same tension. But it does not generate a BPS network, since any
two adjacent three-junctions must have different winding to accomodate
the three vacuum states in the network, and this would imply that
each side of the hexagonal cell must be seen as BPS and anti-BPS,
simultaneously, frustrating and destabilizing the
network -- see Refs.~{\cite{gto99,cht99,saf99}}.

\begin{figure} 
\centerline{\psfig{figure=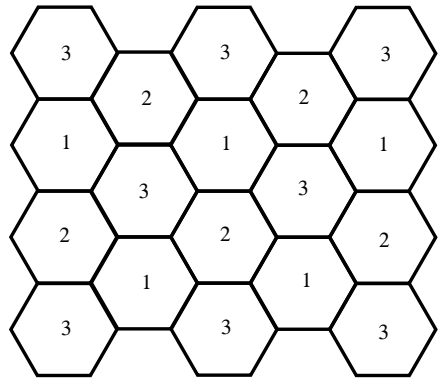,height=5.0cm}} 
\vspace{1.0cm}
\caption{The regular hexagonal network of non-BPS defects
given by the $Z_3$ model defined by the potential
of Eq.~(\ref{p+}).}
\end{figure}

We consider another model, investigated in Ref.~{\cite{bbr99a}}. 
This model is defined by no superpotential, and the argument above 
is no longer valid. The $Z_3$ symmetry seems to allow the presence
of a network of defects, also in the form of regular hexagonal array.
In the thin wall approximation, the hexagonal network that appears
in this case is interesting because the solutions are explicitly known,
form straight-line segments joining the vacuum states, and have the same
energy or tension, $(9/4)\sqrt{27/8}\,|\lambda|$. The $Z_3$ symmetry that
appears in the vacuum sectors is also effective in the sectors of the
topological defects. The inequality $t_{ik}\leq t_{ij}+t_{jk}$ is strictly
valid in this case, ensuring stability of the hexagonal network,
as depicted in FIG.~6 in the thin wall approximation.

\section{Comments and Conclusions} 
\label{sec:comments} 
 
In the present work we have examined several models of coupled scalar fields. 
We have found some explicit solutions, of the BPS and non-BPS type. 
We have also investigated linear stability, checking the behavior of the 
bosonic matter on the background of the BPS and non-BPS defects. The BPS
states are solutions that obey first-order differential equations,
and the non-BPS states are not. The BPS states engender the property of 
having energy evenly distributed in the gradient and potential portions. 
However, in the systems of two coupled fields that we have introduced, 
the non-BPS states also engender the same feature, as we have explicitly 
shown in this work. The point here is that the non-BPS states of the 
systems of coupled fields are BPS states of simpler systems, which are
projections of systems of coupled fields onto a single,
specific field direction. 
 
The first model is the model investigated in Sec.~{\ref{sec:m1}}. It 
presents several properties. In particular, it has a rich BPS sector, 
the sector that connect the minima $(1,0)$ and (-1,0) by one-component 
and two-component BPS states. The two-component state describe a family 
of solutions, parametrized by the ratio between the coupling constants 
that specify the system. These one-component and two-component states 
allow diverse applications, like the applications as bag models in 
Refs.~{\cite{mba96}}, as defects inside defects in
Refs.~{\cite{brs96,bba97,etb98,bbb99}}, and in applications
to condensed matter, to polimeric chains having two relevant degrees
of freedom to represent the chain -- see Refs.~{\cite{brs96e,bve99}}. 
 
In applications in the form of bag models, in addition to the models 
already introduced in Refs.{\cite{rwe75,mon76}} we have other
possibilities, given by the first, second and third models introduced in
Sec.~{\ref{sec:models}}. These models differ from the model considered 
in Ref.~{\cite{pet96}}, and we believe that this new direction 
may shed further light on the issues explored in that work. In particular, 
it seems that our suggestions are easier to examine, and may offer
an analytical understanding of the issues put forward in Ref.~{\cite{pet96}}
under numerical investigations. We recall that the model investigated
in \cite{pet96} is defined by a potential that cannot be reduced to
none of the models considered in the present work.
 
In application to defect inside defect, much work was already done in 
Refs.~{\cite{brs96,bba97,mor98,etb98,bbb99}}. The natural extension 
is now related to issues where one allows the presence of local 
symmetries, as in the work on superconducting 
strings \cite{wit85}. In the first model of Sec~{\ref{sec:models}} 
the $Z_2\times Z_2$ symmetry can be enlarged by enlarging each one 
of the two independent symmetries. As we have already commented
on in Sec.~{\ref{sec:models}}, we can for instance make the $\chi$
field complex, in this case changing the $Z_2\times Z_2$ symmetry to
$Z_2\times U_l(1)$, if one also includes the Abelian gauge field.
This opens new possibilities, providing for instance a natural
scenario to the Callan-Harvey effect \cite{cha85}. Another possibility
includes models that presents the $Z_2\times Z_3$ symmetry, which may
provide scenarios for domain walls that nest planar networks of topological
defects. These issues are presently under consideration, and we hope to
report on them in the near future.

We have also commented on junctions of defects, and the subsequent 
formation of networks of defects. In this case we need to enlarge 
the spatial dimension from $d=1$ to at least the $d=2$ case. 
In the planar case, we have exemplifyed how to get planar solutions 
from the unidimensional kinks of the model. Here we have considered 
three models, the first and the fourth model of Sec.~{\ref{sec:models}},
and another one, firstly considered in Ref.~{\cite{bbr99a}}.
These systems are all different: 
the first model admits a supersymmetric extension of the type $N=1$, 
and seems to admit no stable network of defects, at least classically;
the fourth model presents junctions that break $1/4$ supersymmetry,
but it also seems to admit no stable network of BPS defects. The other model
does not present a supersymmetric extension, but the defect solutions get
even contributions from the gradient and potential portions of the energy, 
and so they behave very much like BPS states. This is crucial for
showing that the fusion of defects is a reaction that occurs exothermically,
and so the three-junction generates a stable network of defects.

\begin{center} 
Acknowledgments 
\end{center} 
We thank A.L.L. Guedes, A.S. In\'acio, J.R.S. Nascimento, R.F. Ribeiro,
and J. Zanelli for discussions. We also thank CAPES, CNPq,
and PRONEX for partial support.


\end{document}